\documentclass[pre,twocolumn,fleqn,showpacs,showkeys]{revtex4-1}

\usepackage{graphicx}
\usepackage{amssymb}
\usepackage{amsmath}
\usepackage{bm}
\usepackage{textcomp}
\begin{document}
\setcounter{page}{0}
\title{Agent-Based Simulation of the Two-Dimensional Patlak-Keller-Segel Model}
\author{Gyu Ho \surname{Bae}}
\author{Seung Ki \surname{Baek}}
\email[]{seungki@pknu.ac.kr}
\thanks{Fax: +82-51-629-5549}
\affiliation{Department of Physics, Pukyong National University, Busan 48513,
Korea}

\date[]{}

\begin{abstract}
The Patlak-Keller-Segel equation describes the chemotactic interactions of small
organisms in the continuum limit, and a singular peak appears through
spontaneous aggregation when the total mass of the organisms exceeds a critical
value. To deal with this singular behavior numerically, we propose an
agent-based simulation method in which both the organisms and the chemicals are
represented as particles. Our numerical estimates for the threshold behavior are
consistent with the analytic predictions.
\end{abstract}


\maketitle
\thispagestyle{empty}

\section{Introduction}

Numerical approaches to a system of partial differential equations take up an
important part of computational physics due to the wide range of applications
from cosmology to quantitative finance. A variety of algorithms, such as those
derived from the finite-difference and the finite-element
methods~\cite{newman2013computational,logan2007first,baek2017meshfree}, most of
which are based on the differentiability of the solution, have been developed.
However, as one
sees from the inviscid Burger's equation~\cite{whitham2011linear}, some systems
are known to develop singular behavior, which undermines the foundation of the
existing numerical methods.
Another example is the the Patlak-Keller-Segel (PKS) model that has been devised
to describe chemotaxis, i.e., the movement of organisms in reaction to chemical
stimulus~\cite{patlak1953random,keller1970initiation,horstmann20031970,*horstmann20041970}.
As will be explained below, as the total mass of the organisms exceeds a certain
threshold, the two-dimensional PKS model undergoes an aggregation transition
by amplifying perturbations from a homogeneous solution, and all the mass
eventually condenses into a single point. In other words, the mass
density field develops a singularity at the transition, so the PKS model
also poses a challenging problem in applying the existing numerical
methods based on
differentiability~\cite{epshteyn2009discontinuous,*epshteyn2009fully,*epshteyn2008new,saito2007conservative,strehl2010flux}.
As an alternative, a proposal is to simulate organisms as particles
whereas the chemical density is still represented by a continuous
field~\cite{fatkullin2012study}, whose theoretical counterpart would be the
many-body theory developed in Ref.~\onlinecite{newman2004many}.

The PKS model is worth investigating in its own right as a reaction-diffusion
system for describing collective
patterns~\cite{vicsek1995novel,*czirok1996formation,*vicsek2012collective}. Some
examples include birds
flying in a flock, a school of fish, swarms of insects, and ants forming a
foraging path~\cite{holldobler1990ants,sempo2006spatial,oettler2013fermat}.
An interesting point about the previous examples is that those
patterns are self-organized in the absence of a centralized leader in the
community. Understanding the mechanism of such self-organization is both
theoretically and practically important in biologically motivated fields of
engineering, from biomechanics to swarm
intelligence~\cite{deneubourg1991dynamics,dorigo2004ant,jafar2010ant,tennenbaum2015mechanics}.

In this work, we propose analyzing the PKS model numerically
in a fully agent-based fashion, i.e., treating both the organisms and the
pheromones
as particles. We will show that the numerical result is, indeed, consistent with
the analytic prediction obtained in the continuum limit. The result will also
be discussed in comparison with the semi-agent-based approach proposed in
Ref.~\onlinecite{fatkullin2012study}.
This work is organized as follows: In the next section, we introduce some basic
features of the PKS model. Section~\ref{sec:numeric} explains our
simulation method and presents the results. Discussing implications of
the numerical results, we summarize this work in Section~\ref{sec:summary}.

\section{Model}

\subsection{Patlak-Keller-Segel model}
Here, we will explain the basic idea of the PKS model, following
Refs.~\onlinecite{nanjundiah1973chemotaxis} and
\onlinecite{childress1981nonlinear}. Let $\rho$ denote the number density of
organisms confined in a $d$-dimensional domain called $\Omega$. The total
number of organisms is a conserved quantity, so the variable $\rho$
satisfies the continuity equation:
\begin{equation}
\frac{\partial \rho}{\partial t} + \nabla \cdot \mathbf{j} = 0.
\end{equation}
The current $\mathbf{j}$ has two contributions,
one from the chemical gradient and the other from the
random motions of the organisms. The former is proportional to $\rho \nabla c$,
where $c$ is the concentration of the chemical, and the latter is proportional
to $\nabla \rho$ as in the diffusion equation. Thus, we can write the total
current as
\begin{equation}
\mathbf{j} = \chi_0 \rho \nabla c - \mu_0 \nabla \rho,
\end{equation}
where $\chi_0$ and $\mu_0$ are positive constants denoting the organism's
chemotactic sensitivity and the diffusion coefficient, respectively.
The density of the chemical changes through three processes: Namely, after
having been deposited by the organisms, it diffuses to neighboring sites and
decays
stochastically with a certain rate by chemical degradation. Combining these
three processes, we can write the following equation:
\begin{equation}
\tau \frac{\partial c}{\partial t} = f_0 \rho + \nu_0 \nabla^2 c - g_0 c,
\end{equation}
where $\tau$ is a dimensionless parameter to set the relative time scale of the
chemical dynamics with respect to that of the organism's dynamics, and $f_0$,
$g_0$, and $\nu_0$ are positive constants to denote the specific rates of
pheromone deposition, diffusion, and degradation, respectively. To sum up, the
PKS model can be described as follows:
\begin{subequations}
\label{eq:ks}
\begin{align}
\frac{\partial \rho}{\partial t} &= - \chi_0 \nabla \cdot (\rho \nabla c) +
\mu_0 \nabla^2 \rho, \label{eq:ks1}\\
\tau \frac{\partial c}{\partial t} &= f_0 \rho + \nu_0 \nabla^2 c - g_0 c.
\label{eq:ks2}
\end{align}
\end{subequations}
We impose the Neumann boundary conditions $\nabla \rho \cdot \hat{n} = \nabla c
\cdot \hat{n} = 0$ at the boundary $\partial \Omega$, where $\hat{n}$ denotes
the normal vector to $\partial \Omega$. This means that the normal component of
the
density gradient vanishes everywhere at the boundary, so that the flux across
the boundary is zero. Due to the boundary conditions, the
total mass $M \equiv \int_\Omega \rho ~d\mathbf{r}$ is conserved, where
$\mathbf{r}$ denotes the $d$-dimensional position vector.

\subsection{Power counting}
By using a power counting argument, one can argue that the aggregation
phenomenon described by Eq.~(\ref{eq:ks}) has a critical dimension $d_c = 2$,
below which the population cannot spontaneously collapse into a single
point~\cite{childress1981nonlinear}.
Suppose that a $d$-dimensional aggregate has
a linear size denoted by $\xi$. The number density is then $\rho \sim \xi^{-d}$,
and the differential operator has $\nabla \sim \xi^{-1}$. The deposition of
the chemical is proportional to $\rho$; hence, $\partial c / \partial t \sim
\xi^{-d}$. If $c$ is on the order of $\xi^{-\alpha}$ and the dynamics has a
time scale of $t \sim \xi^{\beta}$, then
\begin{equation}
\alpha + \beta = d.
\label{eq:ab}
\end{equation}
The left-hand side of Eq.~(\ref{eq:ks1}) is, thus, on the order of
$\xi^{-(d+\beta)}$ whereas the two terms on the right-hand side (RHS) scale as
$\xi^{-(2+d+\alpha)}$ and $\xi^{-(2+d)}$, respectively. Note that the former one
is responsible for aggregation while the latter describes diffusion. We assume
that the dominant process is aggregation so that
\begin{equation}
d+\beta = 2+d+\alpha.
\label{eq:power}
\end{equation}
By solving Eqs.~(\ref{eq:ab}) and (\ref{eq:power}) together, we obtain
\begin{equation}
\left\{
\begin{array}{l}
\alpha = \frac{d}{2} - 1,\\
\beta = \frac{d}{2} + 1.
\end{array}
\right.
\end{equation}
As a consequence, the two terms on the RHS of Eq.~(\ref{eq:ks1})
scale as $\xi^{-(1+3d/2)}$ and $\xi^{-(2+d)}$, respectively. If we consider the
limit of $\xi \rightarrow 0$, the description of aggregation remains
self-consistent when $d > 2$, for which $1+3d/2 > 2+d$.
At the critical dimension $d=2$, the competition between aggregation and
diffusion should be determined by using higher-order terms, and
a threshold of the total mass of organisms for aggregation to take
place, as explained below, turns out to exist.

\subsection{Linear stability analysis in two dimensions}
\label{sec:linear}

Let us consider a homogeneous solution $\rho = \frac{g_0}{f_0} c = \rho_{\rm
const}$ on a square plane $\Omega = [0,L] \times [0,L]$.
We examine the linear stability of the solution when small
perturbations are added, i.e., $\rho(x,y,t) = \rho_{\rm const} + \epsilon_\rho
(x,y,t)$ and $c(x,y,t) = \frac{f_0}{g_0} \rho_{\rm const} + \epsilon_c (x,y,t)$.
Due to the boundary conditions, the perturbations can be written as
\begin{subequations}
\label{eq:pert}
\begin{align}
\epsilon_\rho (x,y,t) &= C_1 \cos \frac{m\pi x}{L} \cos \frac{n\pi y}{L}
e^{\eta t},\label{eq:pert1}\\
\epsilon_c (x,y,t) &= C_2 \cos \frac{m\pi x}{L} \cos \frac{n\pi y}{L}
e^{\eta t},\label{eq:pert2}
\end{align}
\end{subequations}
with integers $m$ and $n$, from which we obtain
\begin{equation}
\eta
\begin{pmatrix}
C_1 \\ C_2
\end{pmatrix}
=
\begin{pmatrix}
-k^2 \mu_0 & k^2\chi_0 \rho_{\rm const}\\
f_0 & -g_0 - k^2\nu_0
\end{pmatrix}
\begin{pmatrix}
C_1 \\ C_2
\end{pmatrix},
\label{eq:matrix}
\end{equation}
with $k \equiv (\pi/L) \sqrt{m^2 + n^2}$.
For such perturbations to grow in time with $\eta > 0$, one needs the following
inequality:
\begin{equation}
\mu_0 \left[ g_0 + \pi^2 \left( m^2 + n^2 \right) \nu_0 / L^2
\right] - f_0 \chi_0 \rho_{\rm const} < 0.
\end{equation}
The resulting mass of the organisms is
\begin{eqnarray}
M &=& \rho_{\rm const} L^2 \\
&>& \frac{\mu_0 g_0}{f_0 \chi_0} L^2 + \pi^2 \left( m^2 + n^2 \right)
\frac{\mu_0 \nu_0}{f_0 \chi_0}\\
&\approx& \pi^2 \left( m^2 + n^2 \right) \lambda,
\label{eq:mc}
\end{eqnarray}
where $g_0$ is assumed to be negligibly small, and we have defined
\begin{equation}
\lambda \equiv \mu_0 \nu_0 / (f_0 \chi_0).
\label{eq:lambda}
\end{equation}
The mass for the lowest order perturbation to make a homogeneous
solution become unstable is, therefore,
\begin{equation}
M_c = \pi^2 \lambda.
\label{eq:mclowest}
\end{equation}

\subsection{Bistability}
\label{sec:bistable}

Let us now consider an inhomogeneous
stationary solution of Eqs.~(\ref{eq:ks1}) and (\ref{eq:ks2}) for
$d=2$~\cite{childress1981nonlinear}.
For the sake of analytic tractability,
we assume that the aggregate is formed in the middle of a large two-dimensional
plane. The system may, thus, be approximated as having radial
symmetry so that the solution has only an $r$-dependence in polar
coordinates. In a stationary state, the boundary conditions imply that
$\mathbf{j} = 0$ everywhere. Then, we have
\begin{equation}
\chi_0 \rho \nabla c - \mu_0 \nabla \rho = 0,
\end{equation}
for which the solution is
\begin{equation}
\rho = R \exp(\tilde{c}),
\label{eq:rho_exp}
\end{equation}
where $R$ is a positive constant and $\tilde{c} \equiv (\chi_0 / \mu_0) c$.
Substituting Eq.~(\ref{eq:rho_exp}) into Eq.~(\ref{eq:ks2}), we obtain
\begin{equation}
0 = \lambda^{-1} R \exp(\tilde{c}) + \nabla^2 \tilde{c},
\label{eq:stationary}
\end{equation}
where we have discarded the decay term by taking the small-$g_0$ limit.
This is Liouville's
equation, and it has a solution of the following form~\cite{walker1915some}:
\begin{equation}
\lambda^{-1} R \exp({\tilde{c}})
(u^2 + v^2 + 1)^2 = 2 \left[ \left( \frac{\partial u}{\partial x} \right)^2 +
\left( \frac{\partial u}{\partial y} \right)^2\right],
\end{equation}
where $F(z) = u(x,y) + iv(x,y)$ is an arbitrary analytic function with $z =
x+iy$. If $u/x = v/y = const.$, we have a radially symmetric
aggregate:
\begin{equation}
\rho(r) = \rho_0 \left( 1 + \frac{\rho_0}{8\lambda}r^2 \right)^{-2},
\label{eq:aggregate}
\end{equation}
with $r \equiv \sqrt{x^2 + y^2}$ and $\rho_0 \equiv \rho(r=0)$.
Even if we impose radial symmetry, the equation admits an infinite number of
different solutions, including annular shapes~\cite{walker1915some}. Among
them, Eq.~\eqref{eq:aggregate} has a characteristic length scale $\xi \sim
\int_0^\infty \rho(r) r^2 ~dr / \int_0^\infty \rho(r) r~dr = \pi \sqrt{2\lambda/
\rho_0}$ so that $\rho_0 \propto \xi^{-2}$, in agreement with the above power
counting argument. The total mass of the organisms in this stationary state is
\begin{equation}
M_0 = 2\pi \int_0^\infty \rho(r) r~dr = 8\pi \lambda.
\label{eq:m0}
\end{equation}

If an aggregate is allowed to form at one of the corners of $\Omega$, as is
often the case in numerical simulations, considering the four-fold symmetry,
the mass for a stationary aggregate should be counted as a fourth of
Eq.~\eqref{eq:m0}~\cite{fatkullin2012study}.
We note that the solution becomes exceedingly complicated on a bounded domain
with other boundary conditions (see, e.g., Ref.~\onlinecite{tracy1986real}).
However, if we have a sufficiently large $L$ compared to the width of the
aggregate, we may expect this kind of calculation still to give a reasonable
estimate.
On the other hand, we have seen that
the homogeneous solution is linearly stable up to $M =
\pi^2 \lambda$ [Eq.~\eqref{eq:mclowest}].
This suggests that the system will be bistable when
\begin{equation}
2\pi \lambda \lesssim M \lesssim \pi^2 \lambda.
\label{eq:bistable}
\end{equation}

\subsection{Coarsening of aggregates}
\label{sec:coarsen}

Depending on the initial condition, the system may develop multiple aggregates,
and the question is then how these aggregates interact with one
another~\cite{velazquez2004point1,*velazquez2004point2,dolbeault2009two}.
An instructive assumption would be for
the chemical dynamics to be fast enough to set
$\tau=0$ in Eq.~(\ref{eq:ks2})~\cite{fatkullin2012study}.
We then obtain an inhomogeneous modified Helmholtz equation,
\begin{equation}
\left( \nabla^2 - \kappa^2 \right) c = - \frac{f_0}{\nu_0} \rho.
\end{equation}
where $\kappa^{-1} \equiv \sqrt{\nu_0 / g_0}$ defines a length scale, meaning
how far the chemical travels before decaying at time $\sim g_0^{-1}$.
The formal solution is given by a convolution formula,
\begin{equation}
c(\mathbf{x},t) = -\frac{f_0}{\nu_0} (\mathcal{G} \ast \rho) (\mathbf{x},t)
= -\frac{f_0}{\nu_0} \int d\mathbf{y}~ \mathcal{G} (\mathbf{x} - \mathbf{y})
\rho(\mathbf{y},t),
\label{eq:convolution}
\end{equation}
with a Green's function $\mathcal{G}$. In two dimensions, it takes the following
form:
\begin{equation}
\mathcal{G}(\mathbf{x}) =
-\frac{K_0 (\kappa x)}{2\pi},
\label{eq:G}
\end{equation}
where $K_0$ is the modified Bessel function of the second kind. For a small
argument $z \ll 1$, it can be approximated as
\begin{equation}
K_0 (z) \approx
-\gamma - \ln \left( \frac{z}{2} \right),
\label{eq:smallz}
\end{equation}
where $\gamma \approx 0.5772$ is the
Euler-Mascheroni constant. Such logarithmic behavior at a short
distance is a signature of two-dimensional gravity, and this behavior
crosses over to an exponential decay as the distance exceeds $\kappa^{-1}$.
Substituting Eq.~(\ref{eq:convolution}) into
Eq.~(\ref{eq:ks1}), we get a closed equation for $\rho$,
\begin{equation}
\frac{\partial \rho}{\partial t} = \nabla \cdot \left[ \frac{f_0 \chi_0}{\nu_0}
\rho \nabla (\mathcal{G} \ast \rho) + \mu_0 \nabla \rho \right] = \nabla \cdot
\left[ \rho \nabla \frac{\delta \mathcal{E}}{\delta \rho} \right],
\label{eq:closed}
\end{equation}
with
\begin{equation}
\mathcal{E} [\rho(\mathbf{x})]
\equiv \mu_0 \int \rho(\mathbf{x}) \ln \frac{\rho(\mathbf{x})}{\rho_{\rm ref}}
d\mathbf{x}
+ \frac{f_0 \chi_0}{2 \nu_0} \iint \rho(\mathbf{x}) \mathcal{G}(\mathbf{x} -
\mathbf{y}) \rho(\mathbf{y}) d\mathbf{x} d\mathbf{y},
\label{eq:E}
\end{equation}
where $\rho_{\rm ref}$ is a constant to make the argument of the logarithm
dimensionless.

Equation~(\ref{eq:closed}) describes relaxational dynamics, in which the
functional $\mathcal{E}$ is non-increasing:
\begin{eqnarray}
\frac{d\mathcal{E}}{dt} &=& \int \frac{\partial \rho(\mathbf{x})}{\partial t}
\frac{\delta \mathcal{E}}{\delta \rho} d\mathbf{x}\\
&=& \int \nabla \cdot \left( \rho \nabla \frac{\delta \mathcal{E}}{\delta \rho}
\right) \frac{\delta \mathcal{E}}{\delta \rho} d\mathbf{x}\\
&=& - \int \left| \nabla \frac{\delta \mathcal{E}}{\delta \rho} \right|^2
\rho(\mathbf{x}) d\mathbf{x} \le 0,
\end{eqnarray}
where we have used integration by parts in deriving the last line. This
inequality shows that $\mathcal{E}$ is a Lyapunov functional (see also
Refs.~\onlinecite{biler1998local,calvez2008parabolic,baek2017free,*bae2019discontinuous}).
If we interpret Eq.~(\ref{eq:E}) as the free energy in the limit of fast
pheromone
dynamics, the first and the second terms will correspond to its entropic and
energetic contributions, respectively. The latter part suggests that the
interaction potential between positions $\mathbf{x}$ and $\mathbf{y}$ is given
by the Green's function $\mathcal{G} (\mathbf{x} - \mathbf{y})$, which
essentially describes two-dimensional gravity with a finite range of
interaction, $\kappa^{-1}$.

\section{Method and Result}
\label{sec:numeric}

Let us begin this section by reviewing the numerical approach of
Ref.~\onlinecite{fatkullin2012study} in
which organisms are represented as particles whereas the chemical density $c$ is
represented as a field variable defined at each lattice point. This choice may
be justified by the fact that $c$ exhibits a weaker
divergence than $\rho$ [see Eq.~\eqref{eq:rho_exp}]. The number of particles is
$N$, so each particle carries mass $m \equiv M/N$. The position of the $n$th
particle is denoted by $X^{(n)}$, and it is updated by chemotaxis and diffusion
by using the explicit Euler scheme
\begin{equation}
X^{(n)}(t+\Delta t) = X^{(n)}(t) + \chi_0 \nabla c \Delta t + \sqrt{2\mu_0
\Delta t}\mathcal{N}(0,1),
\label{eq:x}
\end{equation}
where $\mathcal{N}(0,1) $ means a random number drawn from the Gaussian
distribution with zero mean and unit variance.
The author of Ref.~\onlinecite{fatkullin2012study} argued that the empirical
probability density for $X^{(n)}$ converges to $\rho$, satisfying
Eq.~\eqref{eq:ks1}, as $N \to \infty$.
Numerically,
the mass density field $\rho$ is reconstructed from the set of $X^{(n)}$'s
through bilinear interpolation. It is then used as an input to solve a system of
coupled linear equations, a lattice-discretized version of Eq.~\eqref{eq:ks2},
by using an implicit method.
As in Section~\ref{sec:linear}, we consider $\Omega = [0,L] \times [0,L]$
throughout this section, and the reflecting boundary condition is
imposed to simulate the Neumann boundaries at the particle level.

The difference of our method lies in the way we deal with the chemical density
field $c$.
Here, the chemicals are also represented as particles,
and their dynamics consists of
three parts: deposition, diffusion, and degradation [Eq.~\eqref{eq:ks2}].
First, an organism particle (OP) deposits a pheromone particle (PP) with
probability $f_0 \Delta t/\tau$ at each time step. Observing that
the deposition rate is proportional to the mass density of the organisms, not to
the number density of OP's, is important. This implies that we need to consider
the dynamics in terms of \emph{mass} so that each PP should also carry the
weight of $m$ just as an OP does. The number of PP's is not conserved
during the simulation. Let $Y^{(n)}$ denote the position of the $n$th PP.
This variable is updated by diffusion, again by using the explicit Euler scheme
\begin{equation}
Y^{(n)}(t+\Delta t) = Y^{(n)}(t) + \sqrt{2\nu_0\Delta t/\tau}\mathcal{N}(0,1).
\label{eq:y}
\end{equation}
Finally, the pheromone particle is eliminated with probability $g_0 \Delta t
/\tau$.

In simulations, we take different numbers of OP's: $N=100$,
$400$, and $1000$. We also need to set the number of grid cells along each
direction, $B$, to reconstruct the density fields $\rho$ and $c$. Our criterion
is that the number of grid cells in $\Omega$ should be smaller than or
comparable to that of the OP's, i.e., $B^2 \lesssim N$.
We check different values of $m$, the mass of each OP, to see how the system's
behavior changes as the total mass $M$ varies while keeping $N$ constant.
For each $M$, we generate six independent samples and run the simulations up to
$t = O(10^4)$.
The initial condition is based on a uniformly random distribution of OP's
with no PP's, and a bias toward the origin has been added as a perturbation,
following Ref.~\onlinecite{fatkullin2012study}.
For the modeling parameters, we choose $\mu_0 = 5\times 10^{-3}$, $\nu_0 = 1$,
$\chi_0 = 10^{-1}$, $g_0 = 10^{-2}$, $f_0 = 1$, and $\tau= 10^{-1}$. This gives
us $\lambda \equiv \mu_0 \nu_0 / (f_0 \chi_0) = 5 \times 10^{-2}$
[Eq.~\eqref{eq:lambda}].
The size of $\Omega$ is given by $L=3.2$, and the time step is $\Delta t =
10^{-3}$.
We use this $\Delta t$ so that the displacement of each
PP in Eq.~\eqref{eq:y} does not exceed the lattice constant $a
\equiv L/B$. A chemotactic interaction can, nevertheless, cause an OP's
displacement to be greater than $a$ with this $\Delta t$. Therefore, we have
additionally implemented the adaptive time-stepping method as proposed in
Ref.~\onlinecite{fatkullin2012study}: If the displacement of the $n$th OP
becomes greater than $a$, we subdivide $\Delta t$ into smaller pieces, e.g., 10
pieces of $\Delta t/10$, and find $X^{(n)} (t+\Delta t)$ by updating
$X^{(n)}(t)$ ten times with this smaller time interval, $\Delta t/10$. Just as
we do for $\rho$, we reconstruct the chemical density field $c$ through
bilinear interpolation of the positions of the PP's, which is then used as an
input to calculate the RHS of Eq.~\eqref{eq:x}.

Our main observable is defined as follows:
\begin{equation}
\Sigma_M \equiv \frac{1}{M^2} \int_\Omega \rho^{2}(\mathbf{r})~ d\mathbf{r},
\end{equation}
where the prefactor is necessary for adjusting the trivial dependence on
$M$ even when the distribution keeps the same shape. The more evenly the
distribution $\rho$ spreads, the smaller the value of $\Sigma_M$ takes.
This observable is related to the participation ratio in the
localization problem~\cite{edwards1972numerical} and is commonly used to measure
inhomogeneity~\cite{lee2016evolution}.
Numerically, the integral is calculated with the Riemann sum over the lattice
points.

\begin{figure}
\includegraphics[width=0.45\textwidth]{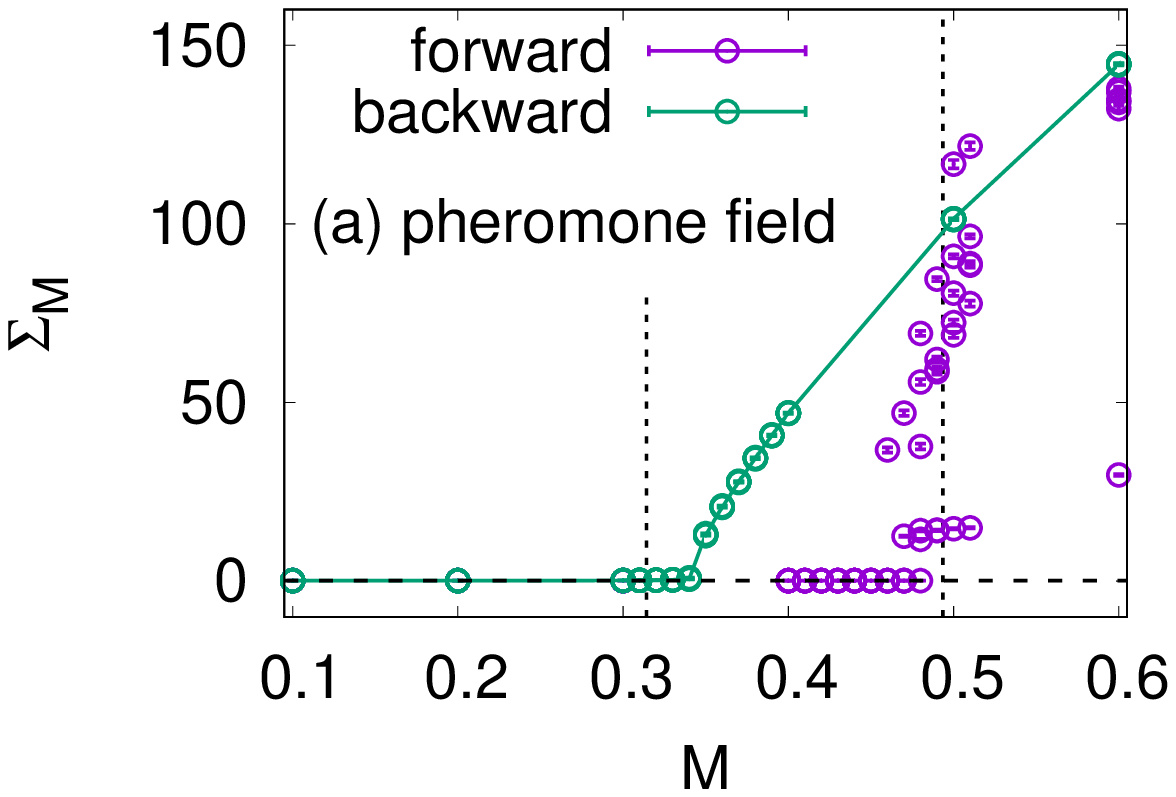}
\includegraphics[width=0.45\textwidth]{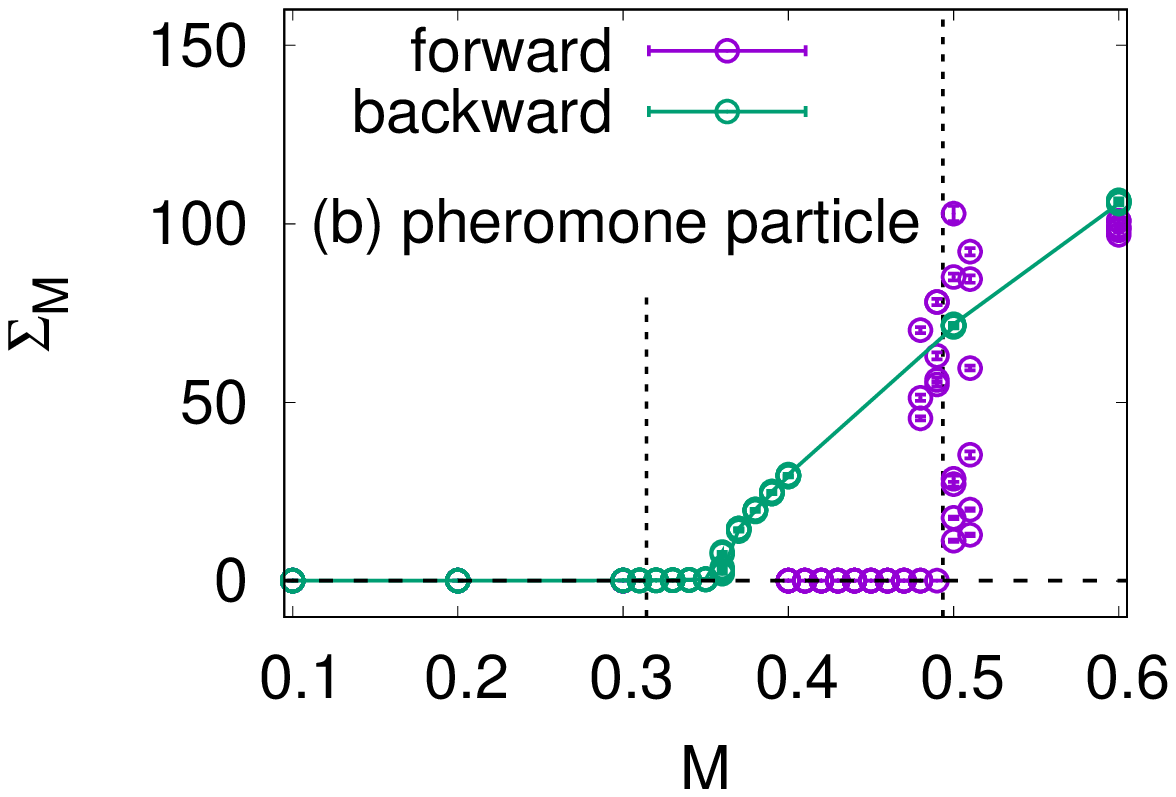}
\caption{$\Sigma_M$ versus $M$ at $t=3\times 10^4$ when $N=10^3$ and $B=32$.
For the modeling parameters, we choose $\mu_0 = 5\times 10^{-3}$, $\nu_0 = 1$,
$\chi_0 = 10^{-1}$, $g_0 = 10^{-2}$, $f_0 = 1$, and $\tau= 10^{-1}$. This gives
us $\lambda \equiv \mu_0 \nu_0 / (f_0 \chi_0) = 5 \times 10^{-2}$
[Eq.~\eqref{eq:lambda}]. The size of $\Omega$ is given by $L=3.2$, and the time
step is $\Delta t = 10^{-3}$.
We take six
independent samples for each $M$. By `forward' and `backward', we mean that $M$
is increased and decreased, respectively, after reaching a steady state for
given $M$. (a) Results from the method in Ref.~\onlinecite{fatkullin2012study}
where pheromones are represented as a field variable $c$ defined at each lattice
point, and (b) those from our method, where pheromones are represented as
particles. The horizontal dashed lines mean $\Sigma_M$ for $N$ randomly
distributed points, and the vertical dotted lines mean the analytic thresholds
in Eq.~\eqref{eq:bistable}, i.e., $2\pi \lambda \approx 0.314$ and
$\pi^2/\lambda \approx 0.493$.
}
\label{fig:n1000}
\end{figure}

\begin{figure}
\includegraphics[width=0.45\textwidth]{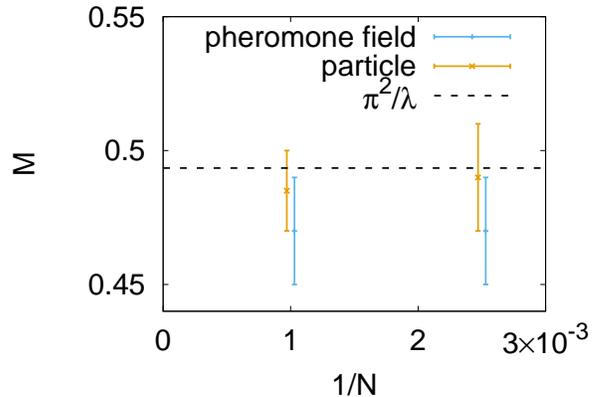}
\caption{Threshold estimates from the two numerical methods:
One is proposed in Ref.~\onlinecite{fatkullin2012study} where pheromones are
represented as field variables, and the other is ours, where pheromones are
represented as particles. The horizontal dashed line is the predicted
upper threshold from the linear stability analysis, $M_c = \pi^2 \lambda \approx
0.493$ [Eq.~\eqref{eq:mclowest}]. We are not showing the results for $N=100$,
where both the methods yield equally imprecise estimates because of large
statistical fluctuations in particle dynamics.}
\label{fig:threshold}
\end{figure}

To check our approach,
we have reproduced the method of Ref.~\onlinecite{fatkullin2012study},
in which only the organisms are represented as particles. The result is shown
in Fig.~\ref{fig:n1000}(a), and it is qualitatively consistent with the
analytically predicted bistable behavior in Eq.~\eqref{eq:bistable}. Based on
this existing method, we have implemented our own, as explained above.
The result is depicted in Fig.~\ref{fig:n1000}(b),
which is also in good agreement with the analytic prediction.
The thresholds can readily be estimated from these numerical results, and
the estimates for the upper threshold, $M_c = \pi^2 \lambda \approx
0.493$, are shown in Fig.~\ref{fig:threshold}.
The size dependence is small, and
no big difference is observed between $N=400$ and $1000$.
Compared to the pheromone-field method in Ref.~\onlinecite{fatkullin2012study},
we see that our pheromone-particle method gives a better estimate for the upper
threshold.

\section{Discussion and Summary}
\label{sec:summary}

In this work, we have proposed solving the PKS model numerically in a fully
agent-based fashion to simulate its singular behavior of aggregation. We have
found that our simulation method successfully reproduces the analytic
predictions. In Eq.~\eqref{eq:bistable}, we have estimated the bistable region.
With our parameter values, the upper threshold is given as $M_c  = \pi^2/\lambda
\approx 0.493$. We have estimated the upper threshold as being as somewhere
between $0.47$ and $0.50$ (Fig.~\ref{fig:threshold}).
We also observe a qualitatively consistent result by using the pheromone-field
method in Ref.~\onlinecite{fatkullin2012study}. However, if we look more
closely, this latter method tends to underestimate the threshold as one sees in
Fig.~\ref{fig:threshold}.
A possible explanation for this deviation could be that the diffusion
coefficient $\mu_0$ is renormalized when the continuum limit is taken for $c$
first~\cite{newman2004many}: The perturbative calculation is based on an
assumption that $\nu_0 \gg \mu_0$, which is valid in our case where $\nu_0 /
\mu_0 = 2 \times 10^2$. The correction terms involve a cutoff length called
$\Lambda$. If we take it as being identical to the lattice constant
$a = 10^{-1}$ for $N=10^3$ and calculate the correction to the second order as
given in Ref.~\onlinecite{newman2004many}, the corresponding threshold value is
lowered to $0.454$. On the other hand, the bare parameter values are preserved
if one takes the limiting process simultaneously for $\rho$ and $c$
~\cite{stevens2000derivation}.

The lower threshold in Eq.~\eqref{eq:bistable} is more difficult to check.
It has been estimated to be $M_0 = 2\pi \lambda \approx 0.314$ from the
stationarity of an aggregate. Our numerical approach shows that the aggregation
persists until the mass decreases to $M=0.36$, but this is still greater than
the analytic prediction. This is presumably because
the calculation in Section~\ref{sec:bistable} is based on an
assumption that $\Omega$ is an unbounded plane $\mathbb{R}^2$. We are
working with a finite square domain throughout this work,
so the threshold can plausibly
become greater than in $\mathbb{R}^2$. The reason is as follows: As
discussed in Section~\ref{sec:coarsen}, our system can be regarded as a
gravitational model, and the method of images under the Neumann boundary
condition suggests that an organism and its image will also interact each other.
In effect, therefore, an organism will experience an attractive interaction
toward the boundary, which is expected to disperse an aggregate. This implies
that the mass required for a stationary aggregate in $\mathbb{R}^2$ may not be
enough when it comes to a finite domain.
If we alternatively look for the minimum mass required for an organism not to
escape from the neighborhood of $\rho(r) = M \delta(r)$, where $\delta$ is the
Dirac delta distribution, the lower threshold decreases further down to $\pi
\lambda \approx 0.157$ (see Ref.~\onlinecite{fatkullin2012study} for calculating
this threshold through a mapping to the Bessel process). Again, this value has
been derived for $\mathbb{R}^2$, and our simulation was unable to detect this
threshold within the bounded region.

In the context of chemotaxis, the fully agent-based method is intuitively
plausible because, in a sense, it brings us back to the starting point of the
PKS model before taking the continuum limit (see, e.g.,
Ref.~\onlinecite{stevens2000derivation} for the derivation of the PKS model as a
limit of a stochastic particle system), but the point is that it tells us how
to analyze a given PDE system while keeping its numerical instability suppressed
in a more general context.
As one can anticipate, this agent-based method is usually less efficient than
the existing method in Ref.~\onlinecite{fatkullin2012study}, especially when the
specific pheromone deposition rate is high. Developing a hybrid method
to combine the advantages of the two methods would be desirable.

\acknowledgments
This work was supported by a research grant from Pukyong National University
(2017).

%
\end{document}